\documentclass{article}
\usepackage{spconf}
\usepackage[utf8]{inputenc} 
\usepackage[T1]{fontenc}    
\usepackage[]{hyperref}       
\usepackage{url}            
\usepackage{booktabs}       
\usepackage{amsfonts}       
\usepackage{nicefrac}       
\usepackage{microtype}      
\usepackage[numbers]{natbib}
\usepackage{textpos}
\usepackage{tabularx,longtable,multirow}
\usepackage[font={small}]{caption}
\usepackage[table]{xcolor}
\usepackage{fncylab} 
\usepackage{fancyhdr} 
\usepackage[english]{babel}
\usepackage{amsmath}
\usepackage{graphicx}
\usepackage{dsfont}
\usepackage{epstopdf} 
\usepackage{array}
\usepackage{latexsym}
\usepackage{enumerate}
\usepackage{lscape}
\usepackage{subcaption}
\usepackage[normalem]{ulem}
\usepackage[title]{appendix}
\usepackage{tikz}
\usetikzlibrary{automata, positioning, arrows}
\usepackage{musicography}
\usepackage{algorithm} 
\usepackage{algcompatible}
\usepackage{amsfonts}
\usepackage{romannum}
\usepackage{ragged2e}
\usepackage{amsmath}

\algnewcommand\INPUT{\item[\textbf{Input:}]}%
\algnewcommand\OUTPUT{\item[\textbf{Output:}]}%

\makeatletter
\newcommand*{\rom}[1]{\expandafter\@slowromancap\romannumeral #1@}
\usetikzlibrary{shapes,decorations,arrows,calc,arrows.meta,fit,positioning}
\tikzset{
    -Latex,auto,node distance =0.5 cm and 0.8 cm,semithick,
    state/.style ={ellipse, draw, minimum width = 0.7 cm},
    point/.style = {circle, draw, inner sep=0.03cm,fill,node contents={}},
    bidirected/.style={Latex-Latex,dashed},
    el/.style = {inner sep=2pt, align=left, sloped}
}

\title{Synthia's Melody: A Benchmark Framework for Unsupervised \\Domain Adaptation in Audio}

\name{Chia-Hsin Lin$^{\star}$\qquad Charles Jones$^{\star}$  \qquad Bj\"{o}rn W. Schuller$^{\star\dagger}$ \qquad Harry Coppock$^{\star}$ }

\address{$^{\star}$GLAM, Imperial College London, UK \\ $^{\dagger}$Chair EIHW, University of Augsburg, Germany\\ \texttt{cynthialin0130@gmail.com,} \\ \{\texttt{charles.jones17, bjoern.schuller, harry.coppock}\}\texttt{@imperial.ac.uk}}
\begin{document}
\ninept
\maketitle
\begin{abstract}
Despite significant advancements in deep learning for vision and natural language, unsupervised domain adaptation in audio remains relatively unexplored. We, in part, attribute this to the lack of an appropriate benchmark dataset. To address this gap, we present \textbf{Synthia's melody}, a novel audio data generation framework capable of simulating an infinite variety of 4-second melodies with user-specified confounding structures characterised by musical keys, timbre, and loudness. Unlike existing datasets collected under observational settings, Synthia's melody is free of unobserved biases, ensuring the reproducibility and comparability of experiments. To showcase its utility, we generate two types of distribution shifts—domain shift and sample selection bias—and evaluate the performance of acoustic deep learning models under these shifts. Our evaluations reveal that Synthia's melody provides a robust testbed for examining the susceptibility of these models to varying levels of distribution shift.
\end{abstract}

\section{Introduction}
While deep learning models achieve impressive performance across domains such as imaging \citep{he2015deep}, text \citep{devlin2019bert}, and audio \citep{hershey2017cnn}, they are prone to learning \textit{shortcuts} -- features not representative of the intended task \cite{Geirhos_2020}. For example, image classifiers trained to recognise animals may instead depend on spuriously correlated background features \cite{Beery_2018_ECCV}, and natural language models may falsely rely on sentiment when predicting review quality \cite{veitchCounterfactualInvarianceSpurious2021a}. Similar effects manifest across many data types and learning tasks, encompassing imaging, text \citep{niven2019probing}, audio \citep{cleverhans}, and reinforcement learning \citep{lapuschkin2019unmasking}. This tendency poses credibility problems for deploying deep learning methods in high-stakes settings. Notably, shortcuts were leveraged heavily during the COVID-19 pandemic response to achieve falsely high diagnostic accuracy for SARSCoV2 infection using patient respiratory audio \cite{coppock2023audiobased, coppock2021covid}.

Although models may not extract the intended features, shortcut learning needs not be an issue if shortcuts are present at both training and deployment time. In practice, this is unlikely to be the case \cite{arjovsky2019invariant}. We may instead view shortcut learning as part of a larger problem of \textit{distribution shift} (or dataset shift), where the joint distribution of inputs and outputs differs between settings \citep{candela2009dataset}. When building models, we usually assume that training and testing data are identically and independently distributed (\textit{i.i.d}) \cite{vapnikOverviewStatisticalLearning1999}. However, shifts between training and testing stages are ubiquitous in practice \citep{alcorn2019strike}. Trained models may not be robust to dataset shifts in unseen environments.

To address the issue, research on domain adaptation methods aims to improve the robustness of models \citep{ganin2016domainadversarial,kim2019learning,schuster2019debiasing,wang2022generalizing}. However, most of the research focuses on the image and text domains, while relatively few address audio. We attribute this to the lack of common benchmark datasets in audio akin to coloured-MNIST \citep{arjovsky2019invariant} in vision and MNLI \citep{williams-etal-2018-broad} in texts. In audio, \cite{shim23b_interspeech} inject biases into observational speech datasets with five data augmentation methods: mp3 
compression, additive white noise, loudness normalisation, non-speech zeroing, and $\mu$-law encoding. Although the results show that models are prone to the injected shortcuts, unobserved artefacts in the original speech data, such as microphone mismatch, could still alter the experimental outcomes. Moreover, the research only considers cases where training data is 100\% or 0\% perturbed, which limits the ability to observe model behaviours with varying shift levels.

\begin{figure}[]
    \centering
    \includegraphics[width=0.48\textwidth, trim={0cm 0cm 0cm 1.4cm}, clip]{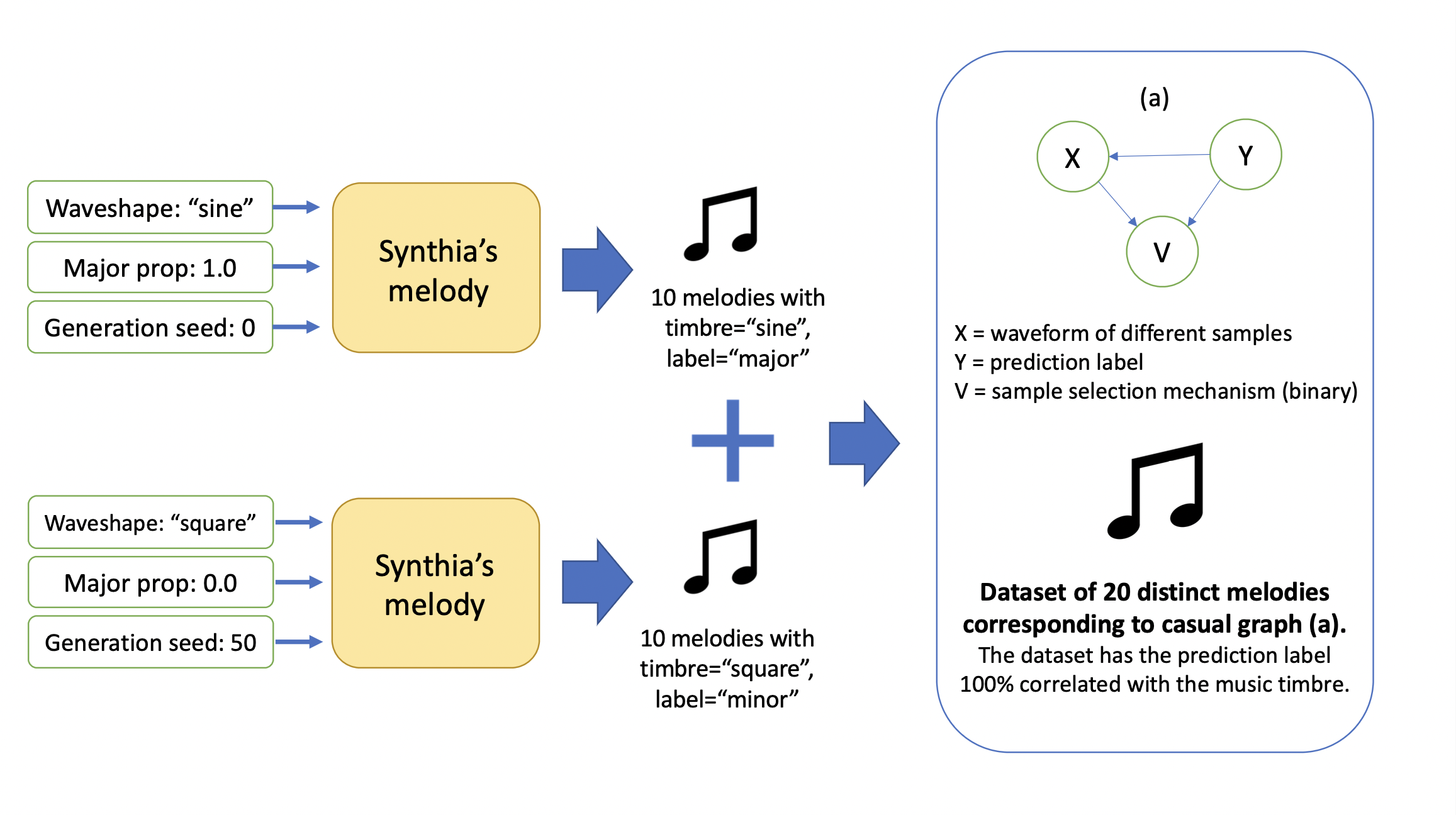}
    \caption{\textbf{Example usage of Synthia's melody}. Here, we see the steps to generate a dataset of 20 samples where the music timbre and prediction label are 100\% correlated. The generated data has a causal structure represented by the causal graph (a). We point the user to the GitHub \texttt{README.md} file for a full set of instructions and use cases.}
    \label{fig:enter-label}
\end{figure}

To foster the corresponding research in audio, we propose a fully synthesised data generation mechanism called \textbf{Synthia's melody}. The mechanism generates 4-second melody samples in the form of \texttt{.wav} file at a sample rate of 16\,000\,Hz for a music key (binary) classification task. The simulated data exhibit several attractive properties. First, it ensures the reproducibility of experiment results, as the data depend only on pre-written scripts in contrast to most audio data collected under observational settings. Second, researchers can generate desired shifts by modifying the mechanism parameters. Third, the generated data are interpretable by humans: researchers can hear the generated melody and feel differences if shifts occur. The example usage is shown in Figure \ref{fig:enter-label}. Code and audio samples are available at \url{https://github.com/cynthpie/Synthia_melody}.

\section{Background}

\noindent \textbf{Domain adaptation} Let $\mathcal{X}$ and $\mathcal{Y}$ denote the input and label space, respectively. We denote the domain $D$ with $D = \{{x}^{(i)}, y^{(i)}\}_{i=1}^{n} \sim p^{D}(x, y)$, where $x\in \mathcal{X}$, $y\in \mathcal{Y}$, and $p^{D}(x, y)$ is the joint distribution of the corresponding random variables $X$, $Y$ that generates $D$. Given that dataset shift exists, the goal of domain adaptation methods is to learn a predictive function $h:\mathcal{X}\rightarrow \mathcal{Y}$ from the source (training) domain $D_{train}$ that minimises the predictive risk $R$ in a similar but unseen domain $D_{test}$, given that $p^{test}(x, y) \neq  p^{train}(x, y)$. The predictive risk function can be written as
\begin{equation}
    R = \min_{h} \mathbb{E}_{(x, y) \in D^{test}}[ \ell(h(x), y)],
\end{equation}
where $\mathbb{E}$ is the expectation and $\ell(\cdot, \cdot)$ is the loss function \footnote{The equation for $R$ represents the theoretical risk we aim to minimise in the target domain $D_{test}$. It is important to note that we cannot directly compute this risk during training given the unsupervised domain adaptation setting.}.\\

\noindent \textbf{Related music theory} We set the generated data in the context of music. Readers unfamiliar with music theory are referred to \cite{roederer2008physics, schmidt2013understanding}. A melody is composed of a sequence of musical tones. Each musical tone has four auditory attributes: pitch, duration, timbre, and loudness. In signal processing, such attributes correspond to frequency, time, waveshape, and amplitude. In Western music, there are 24 music keys; half are major, and the other half are minor. For most people, major keys sound happy and minor keys sound sad. Each music key has seven notes in its corresponding scale. The set of 7 notes is distinct in each of the 24 keys. Certain combinations of the seven notes form chords. Common chords include triads and sevenths, where three notes form triads and sevenths are formed by four. Roman numerals often denote chords. For example, the first triad and fifth seventh in a major key are denoted as \rom{1} and \rom{5}\textsubscript{7}, respectively. We say a melody is in a key if it is composed by the chords in that key.

\section{Method}
To simulate melodies, we need to define four auditory attributes of each musical tone: pitch, duration, timbre, and loudness. \\

\noindent \textbf{Oscillator and ADSR envelope} We determine timbre and loudness with two components: oscillator and ADSR envelope. An oscillator generates waves with a given frequency (pitch) and amplitude (loudness). Oscillators with different waveshapes create different timbres. We use sine, square, sawtooth, and triangle oscillators in this study. 
The ADSR envelope alters the amplitude of waves that oscillators generate. The amplitude change in a melody can be ``stable", ``increase", or ``decrease", where details can be found in the Appendix, amplitude change.\\

\noindent \textbf{Melody generation} Given timbre and loudness defined, we define pitch and duration, or melodies, by using random sampling algorithms. To generate a melody, we randomly draw a music key $K$ among 12 major/minor keys given a label $Y\in \{\text{major}, \text{minor}\}$. We then randomly draw $N$ chords $C_{i=1}^{N}$ in the key $K$, where $N$ is uniformly sampled from some integer set. We consider 10 chords: $C_{\text{major}}=\{\text{\rom{1} \romannum{2}\ \romannum{3}\ \rom{4} \rom{5} \romannum{6}\ \romannum{7}\textsuperscript{o}\ $\text{\romannum{2}}_7$, $\text{\rom{5}}_7$, $\text{\romannum{7}\textsuperscript{\o}}_7$}\}$ and $C_{\text{minor}}=\{\text{\romannum{1}\ \romannum{2}\textsuperscript{o}\ \rom{3}\textsuperscript{+}\ \romannum{4}\ \rom{5} \rom{6}\ \romannum{7}\textsuperscript{o}\ $\text{\romannum{2}\textsuperscript{\o}}_7$, $\text{\rom{5}}_7$, $\text{\romannum{7}\textsuperscript{o}}_7$}\}$ for major and minor samples, respectively. For each chord, we sample their duration $T_{i=1}^{N}$ independently from some continuous distribution. If $\sum{T_{i=1}^{N}}$ is less than 4 seconds, we repeat the melody until the targeted duration is reached. The melody generation algorithm is detailed in the Appendix, algorithm 1. \\


\noindent \textbf{Data generation} We generate 50\,000 melodies with random seeds from 0 to 49\,999 and perform train-val split to obtain 40\,000 training and 10\,000 validation data. We generate 10\,000 melodies with random seeds from 55\,000 to 64\,999. The process is repeated for four timbres: sine, square, sawtooth, and triangle. We fix the amplitude to ``stable" for all samples\footnote{experimenting with using different amplitudes, e.\,g., ``increase'' or ``decrease'' or a custom config left for future work.}. As such, the training, validation, and test sets of the four timbres are acoustically indistinguishable except for their timbres.\\

\noindent \textbf{Shifts considered} We represent distribution shifts with causal graphs \cite{pearl2009causality,castro2020causality}, where the shifts are treated as outcomes of interventions. We focus on anticausal tasks \citep{scholkopf2012causal}, where the input $X$ is caused by the label $Y$, such that $Y\rightarrow X$. We consider two types of shifts: domain shift and sample selection bias \citep{candela2009dataset} detailed in Figure \ref{fig:combined}.

\begin{figure}[h]
    \centering
    
    \begin{subfigure}[b]{.24\textwidth}
        \centering
        \begin{tikzpicture}[
            state/.style={circle, draw, minimum size=2em, inner sep=3pt},
            thick, >={latex}
        ]
        
        \node[state] (Y) at (0,0) {$Y$};
        \node[state] (X_0) [right=of Y] {$X_0$};
        \node[state] (X) [right=of X_0] {$X$};
        \node[state] (f) [above =of X] {$f$};

        \draw[->] (Y) -- (X_0);
        \draw[->] (X_0) -- (X);
        \draw[->] (f) -- (X);
        
        \end{tikzpicture}
        \caption{\textbf{Domain Shift}. The observed covariate $X$ is affected by some mapping $f$ which varies during training and testing. The goal of the classifier is to learn $p(y|x_0)$ via $X$, given that $X_0$ is unobserved.}
        \label{fig:domain_shift_sub}
    \end{subfigure}%
     \hfill
    \begin{subfigure}[b]{.22\textwidth}
        \centering
        \begin{tikzpicture}[
            state/.style={circle, draw, minimum size=2em, inner sep=3pt},
            thick, >={latex}
        ]
        
        \node[state] (Y) at (0,0) {$Y$};
        \node[state] (X) [right=of Y] {$X$};
        \node[state] (V) [below right=of Y, xshift=-0.5cm, yshift=0cm] {$V$};

        \draw[->] (Y) -- (X);
        \draw[->] (Y) -- (V);
        \draw[->] (X) -- (V);
        
        \end{tikzpicture}
        \caption{\textbf{Sample Selection Bias}. The selection process $V\in\{0,1\}$ depends on both the input $X$ and label $Y$. The dependency of $V$ on $X$ and $Y$ varies between training and testing time.}
        \label{fig:samp_selec_sub}
    \end{subfigure}
    
    \caption{The two types of shift considered in this study.}
    \label{fig:combined}
\end{figure}

Domain shift (Figure \ref{fig:domain_shift_sub}) refers to cases when the observed covariate $X$ is affected by some mapping $f$, which varies across training and testing. Given varying $f$, the learnt distribution $p^{train}(y|x)$ has no guarantee to be the same as $p^{test}(y|x)$. The goal of a conditional classifier is to learn the domain-invariant conditional distribution $p(y|x_0)$ via $X$ given $X_0$ unobserved.

Sample selection bias (Figure \ref{fig:samp_selec_sub}) refers to cases when the sample section process $V$ depends on both $X$ and $Y$, and the dependency varies between training and testing time. We denote the selection process with a binary variable $V$, where the event $V=1$ indicates that the sample is being selected and $V=0$, otherwise. Given the varying dependency of $V$, the learnt distribution in training $p^{train}(y|x)$ may not be applicable to that of testing time $p^{test}(y|x)$. \\

\noindent \textbf{Shift construction} We use timbre to construct two types of dataset shifts. For domain shift, we represent two domains by samples generated by sine and square waves. We consider 12 shift levels by gradually replacing the sine sample with the square ones in the training data until the number of sine and square samples are equal. For sample selection bias, we construct a biased sample by correlating the timbre with the prediction label. Specifically, we generate major samples with sine waves and minor ones with square waves. We consider 11 shift levels by varying the proportion of biased samples in the training data. We evaluate trained models on three test sets: in-distribution, neutral, and anti-bias sets.  The in-distribution set is the test set that has the same proportion of biased sample as the training set. The neutral set has no correlation between the timbre and the prediction label. The anti-bias set has the same proportion of the biased samples as the training set, while the bias is reverted, e.\,g. major samples in square waves and minor ones in sine waves. The three test sets are built by extracting the required samples from the sine and square test sets to maintain low compute and time costs. \\

\noindent \textbf{Model} We consider two models: baseline and Domain-Adversarial Neural Network (DANN). The baseline model is a SampleCNN\citep{lee2018samplecnn} with 9 Res-2 blocks\citep{kim2019comparison}. We set all kernel sizes, max-pooling sizes to 3 and stride sizes to 1 in all convolutional layers. The DANN model is a SampleCNN trained with the domain adversarial training algorithm developed by \cite{ganin2016domainadversarial}. A DANN consists of a feature encoder, a classifier, and a domain discriminator. The domain discriminator distinguishes samples in sine and square waves. To make the results of the two models comparable, we use 1 Res-2 block in the feature encoder, 8 Res-2 blocks in the label classifier, and 2 Res-2 blocks in the domain discriminator. In this way, the total parameter sizes of the feature encoder and classifier are the same as the baseline model.\\


\section{Result}
The main purpose of Synthia's melody is to provide a tool researchers can use to evaluate audio-based machine learning models in different dataset shift settings. To demonstrate Synthia's melody's utility, we evaluate the baseline and DANN models across varying levels of shift for a range of different shifts. \\

\begin{figure}[]
    \centering
    \includegraphics[width=0.48\textwidth, trim={0cm 0cm 0cm 1.8cm}, clip]{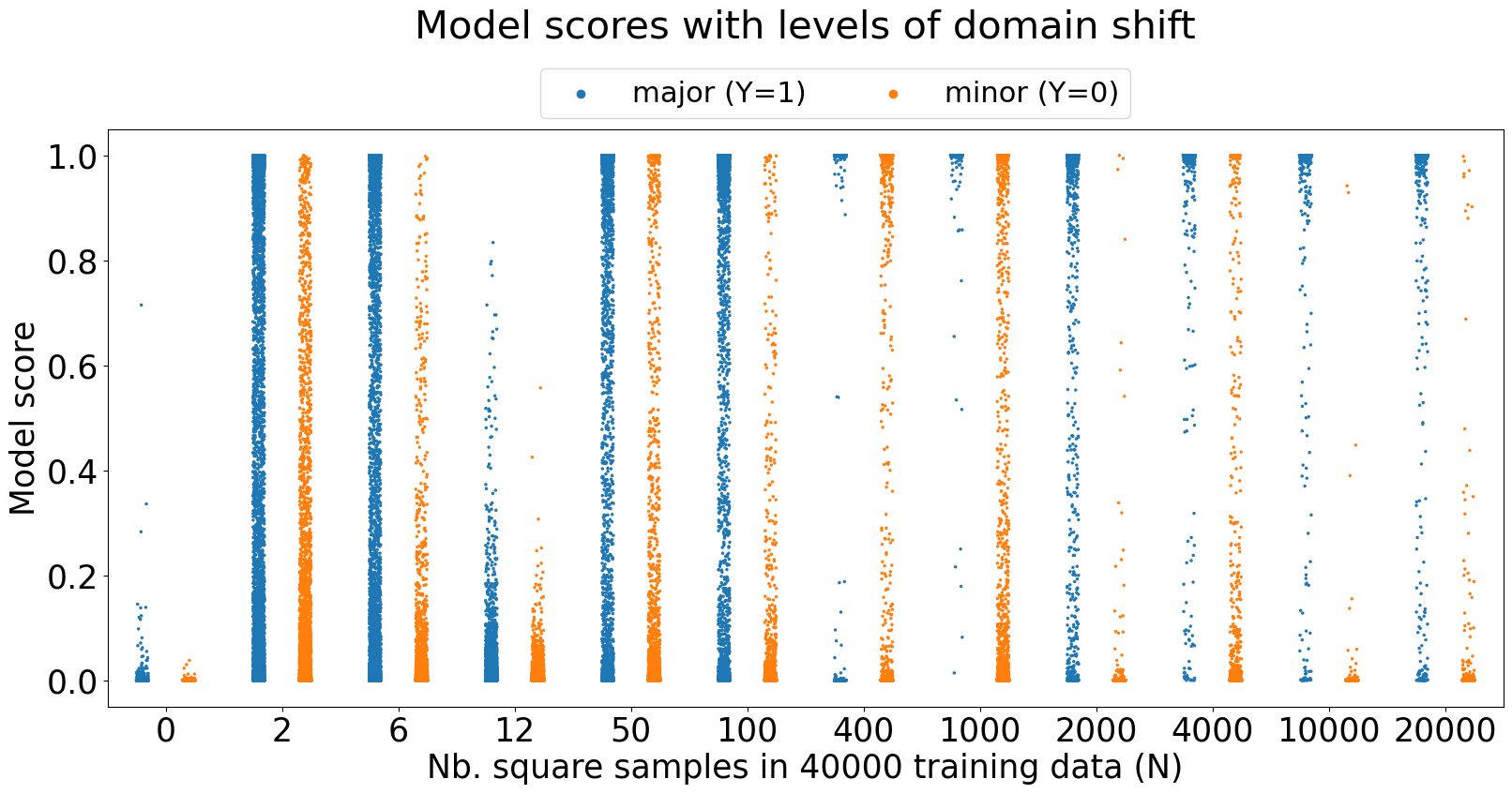}
    \caption{\textbf{Distribution of model score of baseline models on the square test set with varying number of square samples (N) in the training data}. Each point represents the sigmoid prediction of a square test sample. Colours represent the music key, where major samples are expected to have scores close to 1 and minors with 0.}
    \label{fig:ds_score}
\end{figure}

\begin{figure}[h]
    \centering
    \includegraphics[width=\textwidth/2-\textwidth/10]{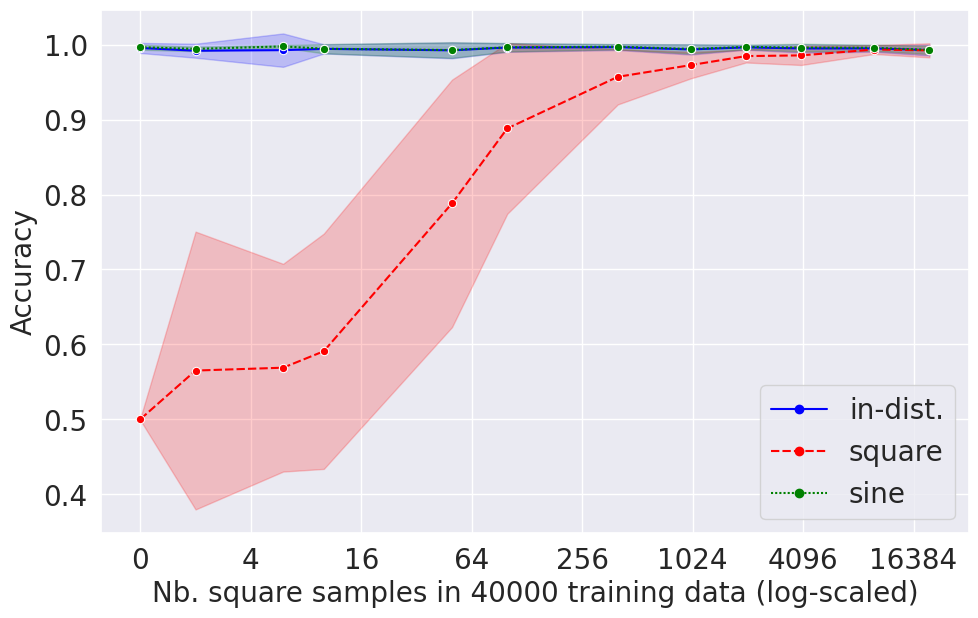}
    \caption{\textbf{Test accuracy of baseline models trained on datasets with 12 levels of domain shift}. The experiment is repeated five times. The line represents the sample mean of the five test accuracies. The 95\% error bands are calculated with $\pm$ 2 standard deviations from the sample mean, assuming that the test accuracy follows a normal distribution.}
    \label{fig:ds_acc}
\end{figure}

\noindent \textbf{Domain Shift}
Figure \ref{fig:ds_score} details the logit output of the baseline model when evaluated on varying levels of domain shift. Here, we see the model performs poorly when no square samples appear in the training set, predicting all square samples as minor. Interestingly, we see a drastic change in behaviour when just two square samples are injected into the training, with the spread of logit scores becoming considerably broader. Inspecting Figure \ref{fig:ds_acc}, which details the corresponding accuracy scores, we see that, despite the behaviour change in logit output, a considerable proportion of square samples are needed before model performance approaches that of an in-distribution test set. Here, we demonstrate how Synthia's Melody can be used to uncover interesting relationships between domain shift, model architecture, and performance. \\

\begin{figure*}[t]
    \centering
    \begin{subfigure}{0.48\linewidth}
        \centering
        \includegraphics[width=\linewidth, trim={0cm 0cm 0cm 1.8cm}, clip]{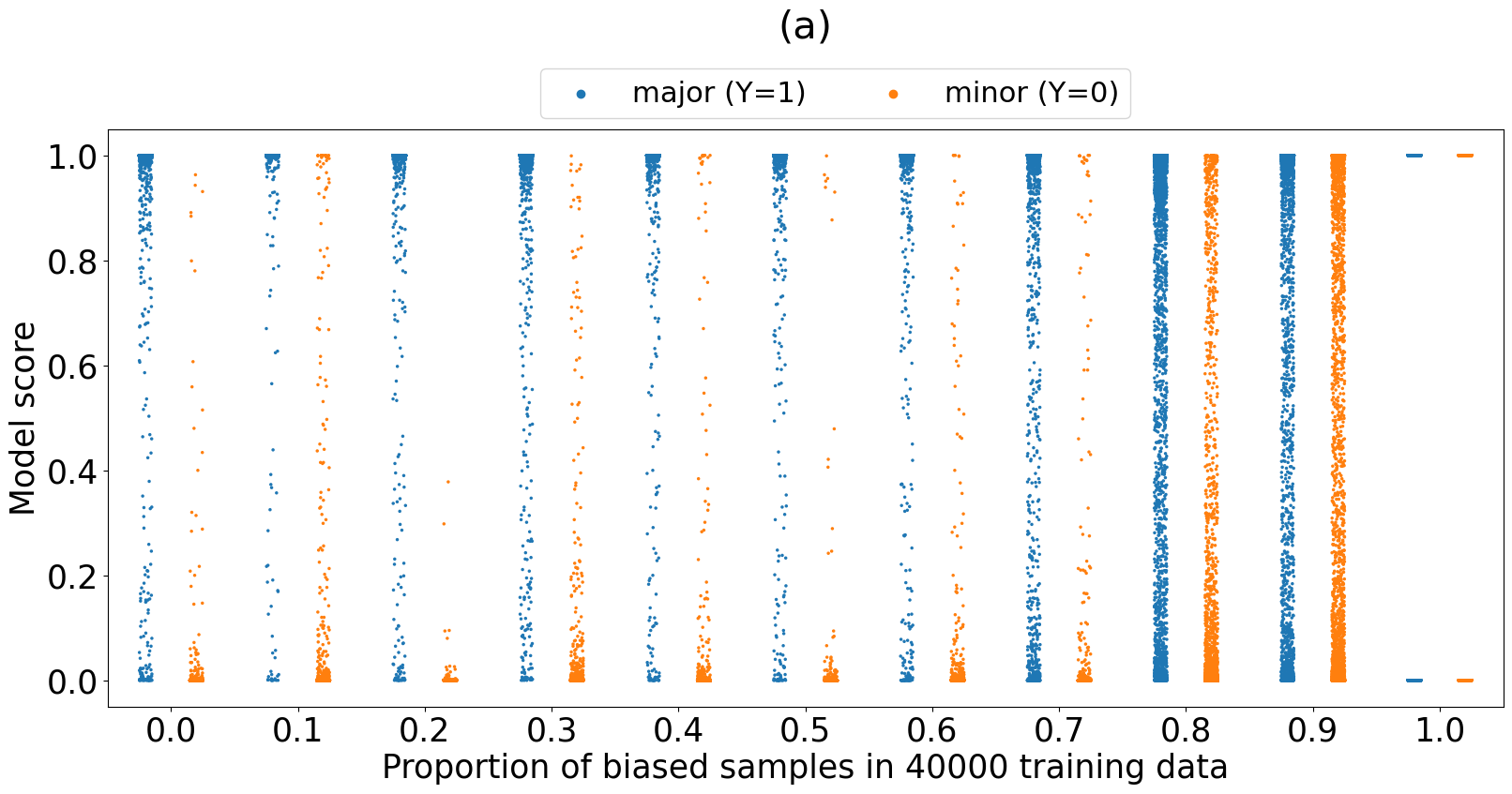}
        \caption{Baseline model scores on the neutral test set with varying proportions of biased sample (shift level) in training data.}
        \label{subfig:sb_score_key}
    \end{subfigure}
    \hfill
    \begin{subfigure}{0.48\linewidth}
        \centering
        \includegraphics[width=\linewidth, trim={0cm 0cm 0cm 1.8cm}, clip]{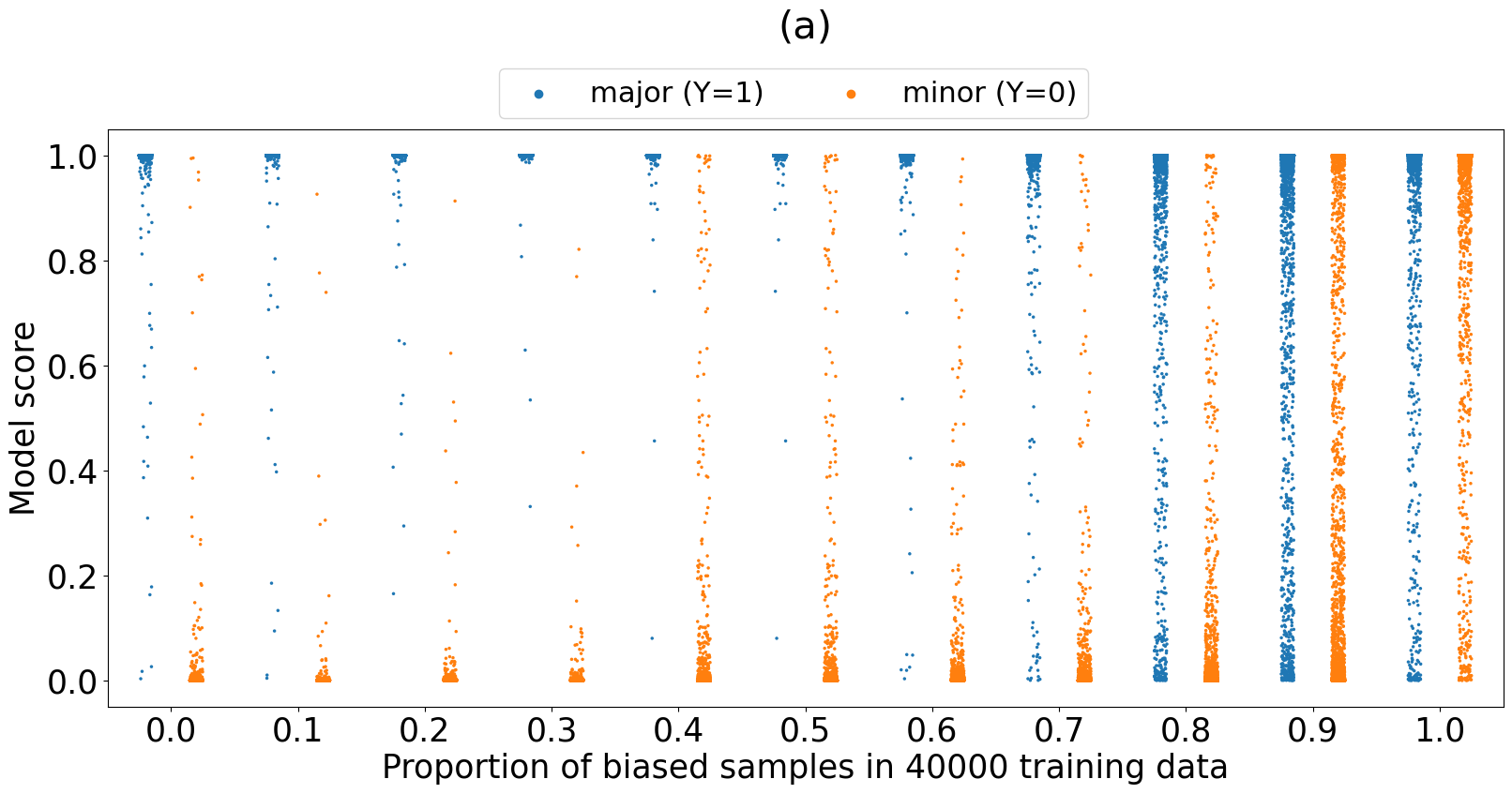}
        \caption{DANN model scores on the neutral test set with varying proportions of biased sample (shift level) in training data.}
        \label{subfig:dann_score_key}
    \end{subfigure}
    \vspace{0.1cm}
    \begin{subfigure}{0.48\linewidth}
        \centering
        \includegraphics[width=\linewidth, trim={0cm 0cm 0cm 1.8cm}, clip]{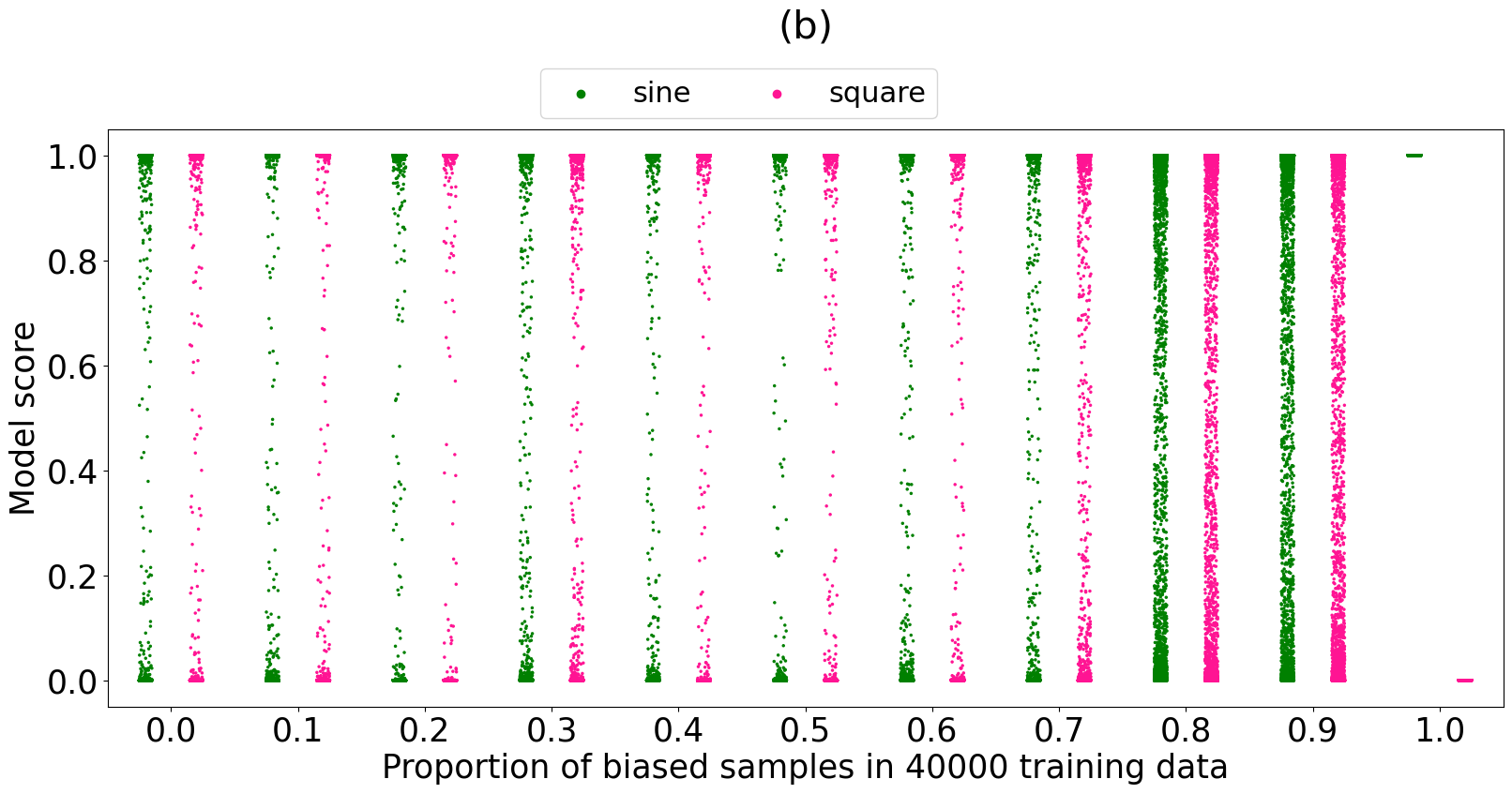}
        \caption{Baseline model scores with wave shape indication.}
        \label{subfig:sb_score_waveshape}
    \end{subfigure}
    \hfill
    \begin{subfigure}{0.48\linewidth}
        \centering
        \includegraphics[width=\linewidth, trim={0cm 0cm 0cm 1.8cm}, clip]{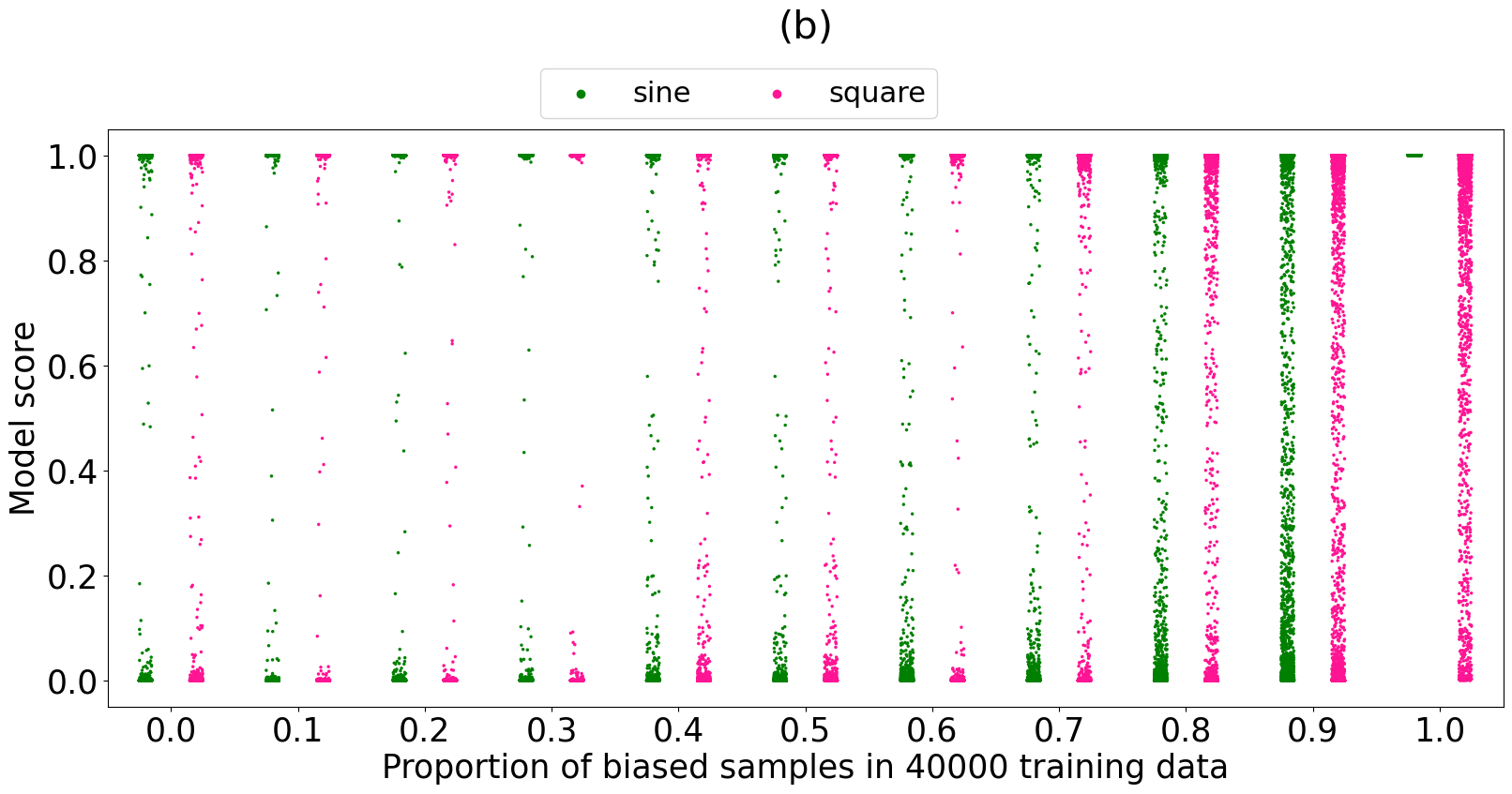}
        \caption{DANN model scores with wave shape indication.}
        \label{subfig:dann_score_waveshape}
    \end{subfigure}
    \caption{\textbf{Comparison of Model and DANN model scores on the neutral test set}. (a) and (c) show the scores of the baseline model, while (b) and (d) show the scores of the DANN model.}
    \label{fig:logit_combined_scores}
\end{figure*}

\noindent \textbf{Sample Selection Bias}
Figure \ref{fig:logit_combined_scores} shows the logit outputs of baseline models on the neutral test set with varying levels of sample selection bias in training data. We examine the model score from two aspects: key (Figure \ref{subfig:sb_score_key}) and the confounding wave shape (Figure \ref{subfig:sb_score_waveshape}). 
At shift levels greater than 0.6, less confident predictions are made, first with major and then with minor classes. When the shift strength is the largest (shift level=1.0), the models do not learn the target task at all, making bias-leading predictions based on wave shapes only -- see Figure \ref{subfig:sb_score_waveshape}.

\begin{figure}[h]
    \centering
    \includegraphics[width=\textwidth/2-\textwidth/10]{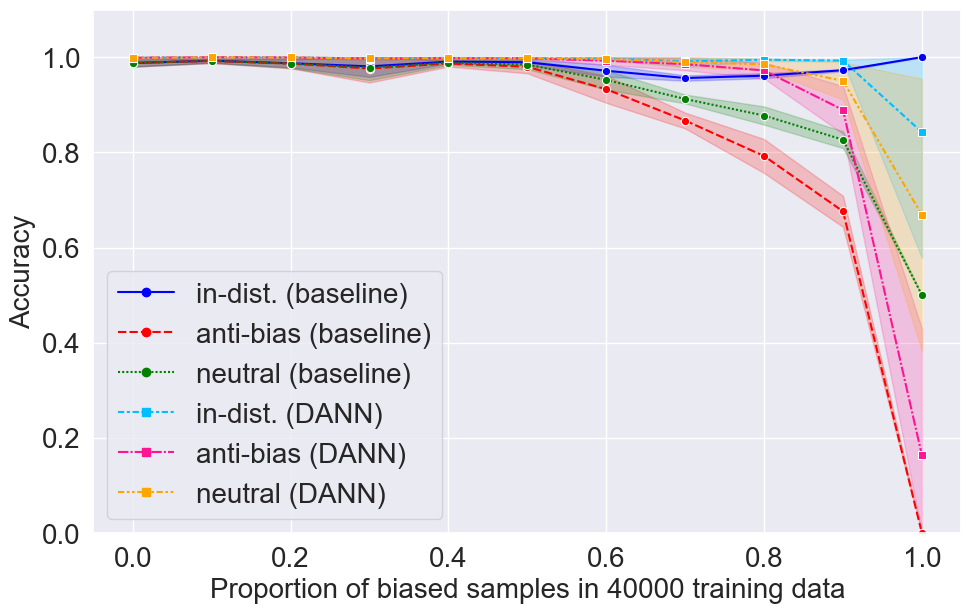}
    \caption{\textbf{Comparison between baseline and DANN models test accuracy across 11 levels of sample selection bias.} The experiment is repeated five times for baseline models and three times for DANN. The line represents the sample mean of the five test accuracies. The bands are calculated with $\pm$ 1 standard deviation from the sample mean, assuming that the test accuracy follows a normal distribution.}
    \label{fig:sb_acc}
\end{figure}

Figure \ref{subfig:dann_score_key} shows the DANN model scores with varying shift levels of sample selection bias injected in training data. The result suggests that DANN is more resistant than the baseline, as scores remain polarised at higher levels of shift than in Figure \ref{subfig:sb_score_key}. It is important to note that at extreme levels of shift, DANN loses 
its robustness, evident in the incorrect prediction of 1 for all sine samples in Figure \ref{subfig:dann_score_waveshape}. 
Figure \ref{fig:sb_acc} compares the test accuracy of the baseline and DANN models at varying levels of sample selection bias. For the baseline model, the three test accuracies diverge at a shift level of 0.7, which is reflected in Figure \ref{subfig:sb_score_key}. We see that as the shift level increases, the model accuracy on the in-distribution set remains high, but the accuracy on neutral and anti-bias sets starts to drop. Specifically, when the shift level equals 1.0, the anti-bias accuracy becomes 0.0, suggesting the baseline model does not learn any signals except the biases. Here, the shortcut is so strong that it acts as a mask, preventing the model from learning any other features despite having ample capacity. This is corroborated by the no-better-than-random neutral test set performance.

\begin{figure}[h]
    \centering
    \includegraphics[width=\textwidth/2-\textwidth/10]{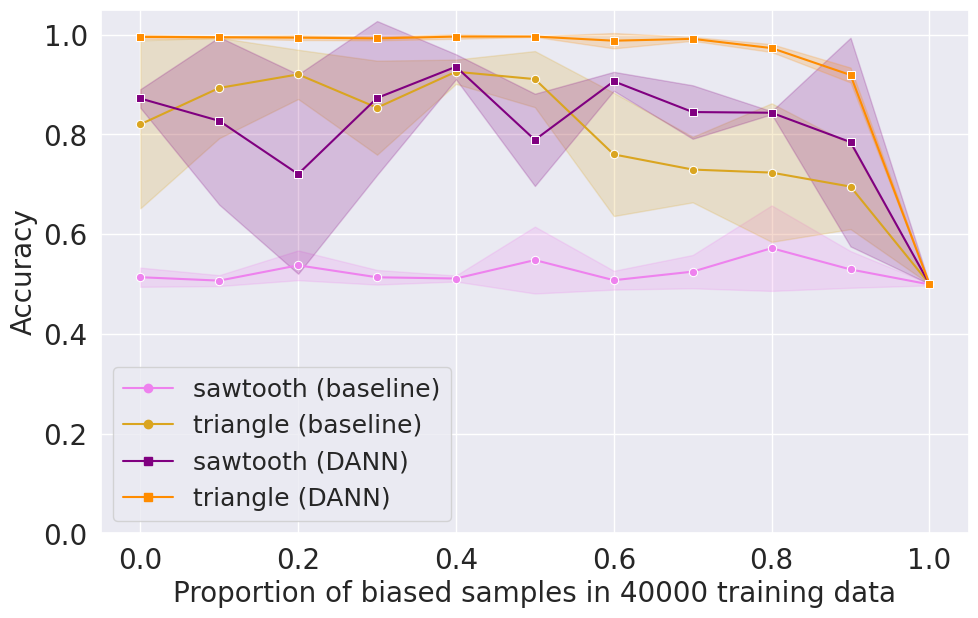}
    \caption{\textbf{Test accuracy of DANN on samples with unseen wave shapes during training}. The x-axis represents the proportion of the biased samples in the sine/square training data. The experiment is repeated five times for baseline models and three times for DANN. The line represents the sample mean and the error bands are calculated with $\pm$ 1 standard deviation from the sample mean.}
    \label{fig:dann_unseen_wave}
\end{figure}

Figure \ref{fig:dann_unseen_wave} shows the baseline model and DANN test accuracy on melodies with unseen wave shapes during training (sawtooth and triangle) as the level of sample selection bias (sine and square) increases in the training set. As shown in Figure \ref{fig:sb_acc}, the test accuracy drops as the shift level increases, identifying a reliance on leveraging sine 
vs square timbre for key prediction. The result shows that DANN is more robust and generalises to unseen wave shapes better than the baseline model. Interestingly, the DANN trained on data with a shift level of 0.4 performs better on unseen wave shapes than ones trained with a lower shift level, such as 0.0 and 0.1. This suggests that injecting a small amount of bias in the training data may actually boost the DANN model's robustness. We hypothesise that a mild amount of biases increases the DANN model's incentive to learn features invariant to the bias feature as the reverted gradients from the discriminator will get larger. There is also the factor of the discriminator acting as a regulariser, injecting noise into the system, through the inverted gradients.

\section{Conclusion}
We presented Synthia's melody, a robust framework for examining the susceptibility of deep audio models to distribution shifts. All melody samples are free of unobserved biases, given their synthetic nature. We detailed novel model behaviour under varying levels of shifts constructed via music timbre. We considered two distribution shifts, domain shift and sample selection bias, and two types of models, a SampleCNN baseline and DANN, where DANN is a SampleCNN trained with a domain-adaptation algorithm. In two types of shift, we showed baseline models make more confident predictions with lower shift levels and notably, in sample selection bias, we showed the injected bias stops the baseline models from learning the target task even if they have enough capacity to do so. The evaluation demonstrates that DANN is more robust to the injected bias and can boost the model's robustness up to, but not including, extreme levels of shift. Synthia's melody provides a robust testbed that allows for reproducible results and serves as an important evaluation framework for the development of future domain adaptation algorithms. Moving forward, the insights gained from Synthia's melody offer vital avenues for enhancing the resilience of deep audio models. 

\bibliographystyle{IEEEtran}
\bibliography{ref.bib}
\clearpage

\pagebreak
\section{Appendix}
\noindent \textbf{Tuning}
We tune all pitches to concert pitch with the reference pitch A4 equals to 440 Hz. Table \ref{tab:concert_pitch} show pitches in the fourth octave and their corresponding frequencies. Pitches in other octaves can be extended by doubling or halving the frequency of the corresponding pitch, e.\,g., A3=220\,Hz and A5=880\,Hz.
\begin{table}[h]
    \centering
    \small
    \renewcommand{\arraystretch}{1.1}
    \begin{tabular}{c|c}
        \textbf{Pitch} &\textbf{Frequency (Hz)}\\
        \hline
        C4 & 261.6256\\
        D$^{\musFlat}$4 & 277.1826\\
        D4 &293.6648\\
        E$^{\musFlat}$4 & 311.1270\\
        E4 & 329.6276\\ 
        F4 & 349.2282\\ 
        G$^{\musFlat}$4 & 369.9944\\ 
        G4 & 391.9954\\ 
        A$^{\musFlat}$4 & 415.3047\\
        A4 & 440.0000\\ 
        B$^{\musFlat}$4 & 466.1638\\ 
        B4 & 493.8833
    \end{tabular}
    \caption{\textbf{Pitches used in Synthia's melody}. This table shows the frequency of the pitches in the 4th octave. Pitches in other octaves can easily be extended by doubling or halving the frequency of the corresponding pitch, e.\,g., A3=220\,Hz and A5=880\,Hz.}
    \label{tab:concert_pitch}
\end{table}

\begin{figure}[h]
    \centering
    \includegraphics[width=\textwidth/4-\textwidth/60]{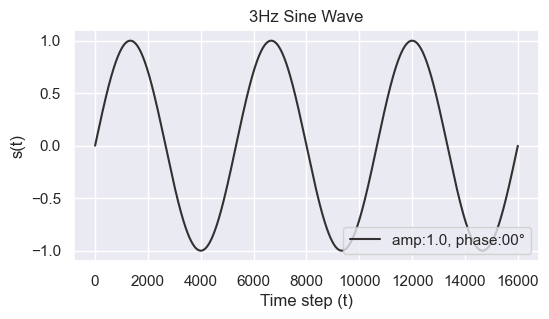}
    \includegraphics[width=\textwidth/4-\textwidth/60]{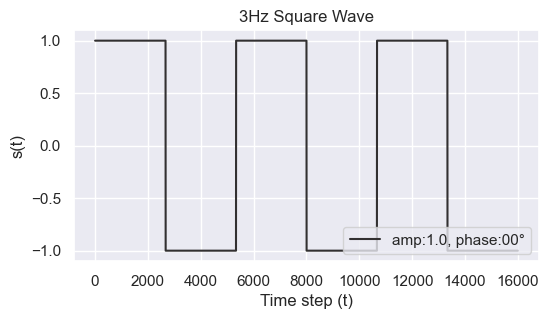}
    \includegraphics[width=\textwidth/4-\textwidth/60]{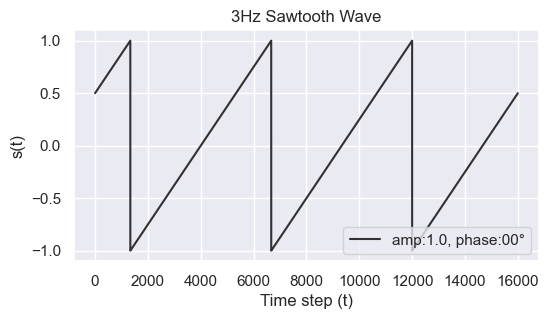}
    \includegraphics[width=\textwidth/4-\textwidth/60]{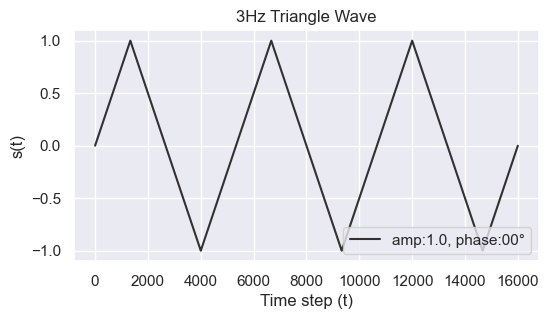}
    \caption{\textbf{Four types of wave shapes}. Each of the waves has a fundamental frequency=3 Hz, amplitude=1.0, and phase=0.0.}
    \label{fig:wave_shapes}
\end{figure}

\noindent \textbf{Waveshape} We consider four types of waveshapes in this study: sine, square, sawtooth, and triangle. The waveshapes are illustrated in Figure \ref{fig:wave_shapes}. Different wave shapes result in different music timbres. Demos are available at \url{https://drive.google.com/drive/folders/13PLu_ZZ7rv9vi5pZWapqebLambOjAElB}. \\

\noindent \textbf{Amplitude change} We use ADSR envelopes to alter the amplitude in melody. An ADSR envelope is defined by the four parameters: attack (A), decay (D), sustain (S), and release (R). We set the amplitude change ``increase" with A=2, D=0.01, S=1, R=0.01; ``decrease" with A=0.01, D=0.01, S=1, D=2, and ``stable" with A=0.01, D=0.01, S=1, D=0.01, where the unit of A, D, R are in seconds, and $S=1$ indicates the maximum volume.\\

\noindent \textbf{Chords considered}
We use major and harmonic minor scales to construct major and minor keys. The corresponding seven basic triads are listed in Table \ref{tab:triad_quality}. In addition to triads, we also considered the sevenths \{\romannum{2}\textsubscript{7}\ \rom{5}\textsubscript{7} \romannum{7}\textsuperscript{\o}\textsubscript{7}\} for major keys and \{\romannum{2} \textsuperscript{\o}\textsubscript{7}\ \rom{5}\textsubscript{7} \romannum{7}\textsuperscript{o}\textsubscript{7}\}
for minor keys. 
\begin{table}[h]
    \centering
    \begin{tabular}{|c|c|c|c|c|c|c|c|}
    \hline
        \textbf{Triads} & \textbf{1} & \textbf{2} & \textbf{3} & \textbf{4}& \textbf{5}& \textbf{6} & \textbf{7}\\
        \hline
        Major key&  \rom{1} & \romannum{2} & \romannum{3} & \rom{4} & \rom{5} & \romannum{6} & \romannum{7}\textsuperscript{o}\\
        Minor key & \romannum{1} & \romannum{2}\textsuperscript{o} & \rom{3}\textsuperscript{+} & \romannum{4} & \rom{5} & \rom{6} & \romannum{7}\textsuperscript{o}\\
        \hline
    \end{tabular}
    \caption{\textbf{Triads considered in major and minor keys}. We denote triads by Roman numerals. The quality of chords are denoted as major (uppercase), minor (lowercase), augmented (+), and diminished (o).}
    \label{tab:triad_quality}
\end{table}

\noindent \textbf{Melody generation algorithm} Algorithm 1 shows the melody generation process used in Synthia's melody. The algorithm can be broken down into eight stages: 1. Draw a music key $K$ (line 1); 2. Construct the scale corresponding to $K$ given the key type $Y\in\{\text{major}, \text{minor}\}$ (line 2); 3. Determine the number of chords present in the melody, denoted as $N$ (line 3); 3. Draw $N$ triads $C_{i=1}^{N}$ from the key $Y$ (line 4); 4. Ensure triads \rom{1}, \rom{4}, \rom{5} (\romannum{1}\, \romannum{4}\, \rom{5} for minor keys) are present (line 6 to 30); 5. Change certain triads to their corresponding sevenths with 0.5 probability (line 31 to 34); 6. Randomly alter the octave of each note in each chord (line 36 to 41); 7. Assign a duration $T$ to each chord to form a melody (line 42 to 44); 8. If the total duration does not reach 4 seconds, repeat the melody until the targeted time is reached (line 45 to 49). In this study, we set $R_f=[130.81, 523.25]$ Hz, i.e. pitches ranged from C2 to C5, $R_n = \{3,4,\dots, 7\}$, and $R_t = [0.2, 0.9]$ seconds.\\

\begin{algorithm*}[]
  \caption{Melody generation process}
  \begin{algorithmic}[1] \label{alg:dgp_2}
    \REQUIRE Frequency range $R_f = [f_{\text{min}}, f_{\text{max}}]$, key type $Y \in \{\text{major}, \text{minor}\}$, number of chords set $R_n = \{n_{\text{lower}},\dots, n_{\text{upper}}\}$, Duration range $R_t=[t_{\text{lower}}, t_{\text{upper}}]$
    \ENSURE One 4-second melody sample
    \STATE Draw a frequency $f_1$ from the 12 frequencies in table \ref{tab:concert_pitch}
    \STATE Construct a set of pitches $S = \langle f_1, f_2, \ldots, f_7 \rangle$ given $f_1$ and $Y$
    \STATE Sample $N$ uniformly from $R_n$
    \STATE Sample $N$ triads from Table \ref{tab:triad_quality} uniformly with replacement given $Y$ to form a set of chord types $S_{\text{type}} = \langle C_{\text{type}}^{(1)}, C_{\text{type}}^{(2)}, \dots, C_{\text{type}}^{(N)} \rangle$, where each $C_{\text{type}}^{(i)}$ is a triad
    \STATE Construct an integer set $L = \{1, 2, \dots, N\}$.
    \IF {$Y=\text{major}$}
        \WHILE{triad \rom{1} and \rom{4} and \rom{5} not in $S_{\text{type}}$}
            \STATE Sample $l$ uniformly from $L$ 
            \IF{triad \rom{1} not in $S_{\text{type}}$}
                \STATE replace $C_{\text{type}}^{(l)}$ with triad \rom{1} 
            \ELSIF{triad \rom{4}not in $S_{\text{type}}$}
                \STATE replace $C_{\text{type}}^{(l)}$ with triad \rom{4}
            \ELSIF{triad \rom{5} not in $S_{\text{type}}$}
                \STATE replace $C_{\text{type}}^{(l)}$ with triad \rom{5}
            \ENDIF
            \STATE Remove $l$ from $L$
        \ENDWHILE
    \ELSE
        \WHILE{triad \romannum{1}\ and \romannum{4}\ and \rom{5} not in $S_{\text{type}}$}
            \STATE Sample $l$ uniformly from $L$ 
            \IF{triad \romannum{1} not in $S_{\text{type}}$}
                \STATE replace $C_{\text{type}}^{(l)}$ with triad \romannum{1}
            \ELSIF{triad \romannum{4}\ not in $S_{\text{type}}$}
                \STATE replace $C_{\text{type}}^{(l)}$ with triad \romannum{4}\
            \ELSIF{triad \rom{5} not in $S_{\text{type}}$}
                \STATE replace $C_{\text{type}}^{(l)}$ with triad \rom{5}
            \ENDIF
            \STATE Remove $l$ from $L$
        \ENDWHILE
    \ENDIF
    \STATE Sample $coin$ uniformly from $\{0,1\}$
    \IF{$coin$=1}
        \STATE Replace all triads \romannum{2}, \rom{5}, \romannum{7}\textsuperscript{o} in $S_{\text{type}}$ with $\text{\romannum{2}}_7$, $\text{\rom{5}}_7$, $\text{\romannum{7}\textsuperscript{\o}}_7$ if $Y=\text{major}$ or $\text{\romannum{2}\textsuperscript{\o}}_7$, $\text{\rom{5}}_7$, $\text{\romannum{7}\textsuperscript{o}}_7$ if $Y=\text{minor}$
    \ENDIF
    \FOR{$i = 1$ to $N$}
        \STATE Construct chord $C_i =\langle p_1, p_2, p_3\rangle$ if triad and $\langle p_1, p_2, p_3, p_4 \rangle$ if seventh corresponds to $C_\text{type}^{(i)}$, where each $p$ is a frequency in $S$
        \FOR{j = 1 to \texttt{length($C_i$)}}
            \STATE Construct a set of integer multiples of $p_j$ in $C_i$, denoted as $O_j=\{\dots, p_j/4, p_j/2,p_j, 2p_j, 4p_j, 8p_j\dots\}\in R_f$. 
            \STATE Sample $o_j$ uniformly from $O_j$
            \STATE Replace $p_j$ with $o_j$
            \ENDFOR
        \STATE Sample $T_i \sim U(t_{\text{lower}}, t_{\text{upper}})$
        \STATE Assign time $T_i$ to chord $C_i = \langle o_1, o_2, o_3\rangle$ if triad or $\langle o_1, o_2, o_3, o_4\rangle$ if seventh
    \ENDFOR
    \STATE Form a melody $m=\langle C_1, C_2, \dots ,C_N\rangle$
    \STATE Calculate $T_m = \sum_{i=1}^{N} T_i$
    \WHILE{$T_m < 4 \text{ seconds}$}
        \STATE Repeat the melody $m$ until $T_m \geq 4$
    \ENDWHILE
  \end{algorithmic}
\end{algorithm*}

\begin{figure}[h]
    \centering
    \includegraphics[width=\textwidth/2-\textwidth/10]{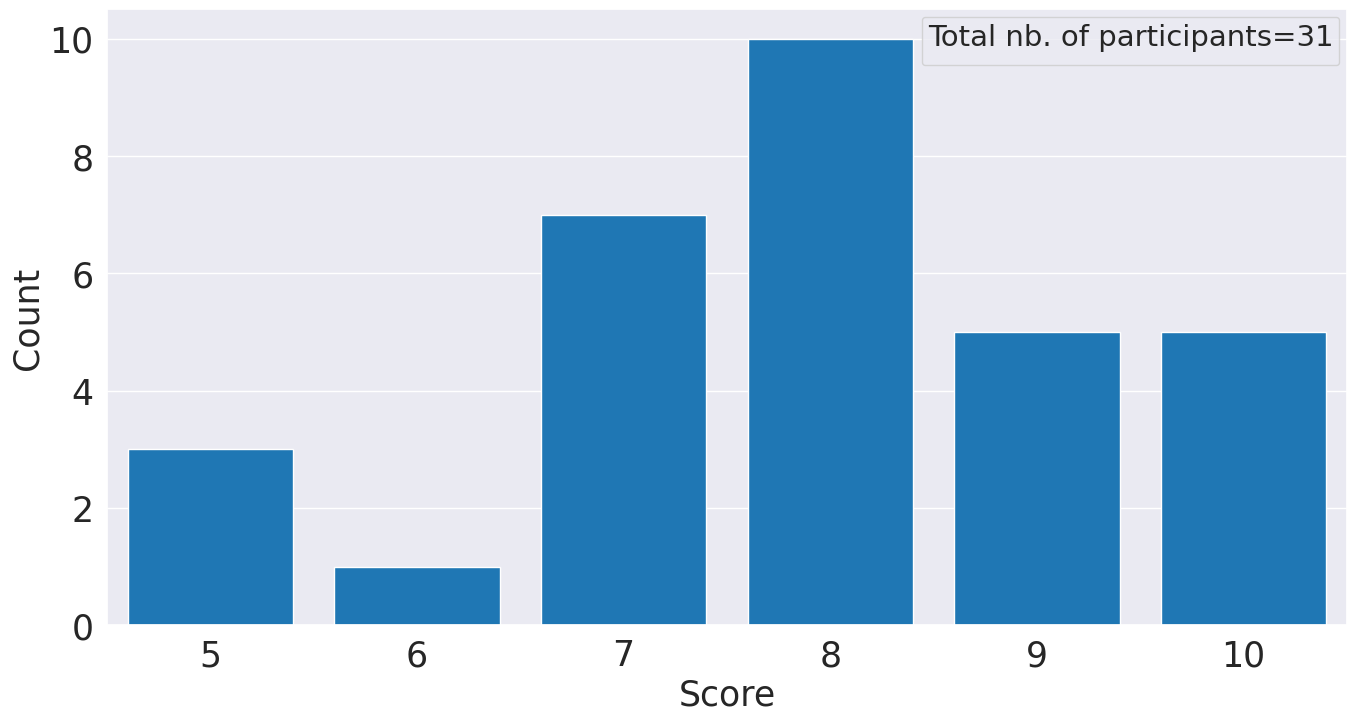}
    \caption{\textbf{Distribution of human evaluation score on 10 melody samples generated by Algorithm 1}. The x-axis represents the number of correctly identified samples out of 10 samples. The y-axis represents the number of participants who achieved the score. The 31 participants achieved an averaged and a median score of 8.}
    \label{fig:dgp_human_resp}
\end{figure}

\noindent \textbf{Human evaluation}
We randomly select 10 samples generated by Algorithm 1 and ask the general audience to perform the music key classification task. We provide one melody demo in a major key and the other one in a minor key at the beginning of the questionnaire for those without professional musical training. The questionnaire is available at \url{https://forms.gle/pan5kaMRxREXtBNK6}. The result is shown in figure \ref{fig:dgp_human_resp}. In the questionnaire, we also asked the participants to provide their professional musical experience in years, and find no significant correlation between higher scores and the musical experience. For example, two of the participants who scored a 10 have no professional musical experience. 

\end{document}